\documentclass[12pt]{iopart}
 \expandafter\let\csname equation*\endcsname\relax
 \expandafter\let\csname endequation*\endcsname\relax 

\usepackage{iopams}
\usepackage{graphicx}
\usepackage{mathtools}
\usepackage{amssymb}
\usepackage{siunitx}
\usepackage{color}
\usepackage[colorlinks=true,citecolor=blue,linkcolor=blue,pdfstartview=FitH,pdfauthor=Gross]{hyperref}
\hypersetup{colorlinks=true}

\expandafter\let\csname equation*\endcsname\relax
\expandafter\let\csname endequation*\endcsname\relax

\begin{document}
\title[Ultrastrong coupling in two-resonator circuit QED]
      {Ultrastrong coupling in two-resonator circuit QED}

\author{A.~Baust$^{1,2,3}$, E.~Hoffmann$^{1,2}$, M.~Haeberlein$^{1,2}$, M.~J.~Schwarz$^{1,2,3}$, P.~Eder$^{1,2,3}$, J.~Goetz$^{1,2}$, F.~Wulschner$^{1,2}$, E.~Xie$^{1,2}$, L.~Zhong$^{1,2,3}$, F.~Quijandria$^{4}$, D.~Zueco$^{4,5}$, J.-J.~Garcia~Ripoll$^{6}$, L.~Garc\'{i}a-\'{A}lvarez$^{7}$, G.~Romero$^{7}$, E.~Solano$^{7,8}$, K.~G.~Fedorov$^{1,2}$, E.~P.~Menzel$^{1,2}$, F.~Deppe$^{1,2,3}$, A.~Marx$^{1}$, R.~Gross$^{1,2,3}$}

\address{$^{1}$Walther-Mei{\ss}ner-Institut, Bayerische Akademie der Wissenschaften, D-85748 Garching, Germany}
\address{$^{2}$Physik-Department, Technische Universit\"{a}t M\"{u}nchen, D-85748 Garching, Germany}
\address{$^{3}$Nanosystems Initiative Munich (NIM), Schellingstra{\ss}e 4, D-80799 M\"{u}nchen, Germany}
\address{$^{4}$Instituto de Ciencia de Materiales de Arag\'{o}n and Departamento de F\'{\i}sica de la Materia Condensada, CSIC-Universidad de Zaragoza, E-50009 Zaragoza, Spain}
\address{$^{5}$Fundaci\'{o}n ARAID, Paseo Mar\'{\i}a Agust\'{\i}n 36, E-50004 Zaragoza, Spain}
\address{$^{6}$Instituto de Fisica Fundamental, IFF-CSIC, Calle Serrano 113b, Madrid E-28006, Spain}
\address{$^{7}$Department of Physical Chemistry, University of the Basque Country UPV/EHU, Apartado 644, E-48080 Bilbao, Spain}
\address{$^{8}$IKERBASQUE, Basque Foundation for Science, Alameda Urquijo 36, E-48011 Bilbao, Spain}

\eads{\mailto{alexander.baust@wmi.badw-muenchen.de}, \mailto{rudolf.gross@wmi.badw-muenchen.de}}


\begin{abstract}
We report on ultrastrong coupling between a superconducting flux qubit and a resonant mode of a system comprised of two superconducting coplanar stripline resonators coupled galvanically to the qubit. With a coupling strength as high as $17\%$ of the mode frequency, exceeding that of previous circuit quantum electrodynamics experiments, we observe a pronounced Bloch-Siegert shift. The spectroscopic response of our multimode system reveals a clear breakdown of the Jaynes-Cummings model. In contrast to earlier experiments, the high coupling strength is achieved without making use of an additional inductance provided by a  Josephson junction.
\end{abstract}


\maketitle

\section{Introduction}

Circuit quantum electrodynamics~(QED)~\cite{Wallraff:2004a} has not only become a versatile toolbox for quantum information processing~\cite{Pechal2014, Chen:2014} and quantum simulation~\cite{Houck2012, Raftery2013, Chen2014weak} but is also a powerful platform for the study of light-matter interaction~\cite{niemczyk_circuit_2010, Forndiaz2010} and fundamental aspects of quantum mechanics~\cite{Menzel2012, Zhong2013, Steffen2013, Roushan2014}. In contrast to the field of cavity QED, where the interaction between a natural atom and the light field confined in a three-dimensional optical cavity is studied, the building blocks of the circuit QED architecture are superconducting quantum bits acting as ‘artificial atoms’ and quasi-onedimensional superconducting transmission line
resonators with resonant frequencies in the microwave regime. Since the mode volumes of the latter are small compared to those of three-dimensional
optical cavities and the dipole moments of the artificial atoms are orders of magnitude larger than those of their natural counterparts, in circuit QED setups the coupling strength between the artificial atom and quantized resonator modes can reach a significant fraction of the system energy. Remarkably,
even the regime of ultrastrong coupling can be reached in superconducting circuits where the Jaynes-Cummings approximation
breaks down~\cite{niemczyk_circuit_2010}. In this situation, the interaction between light and matter can only be described correctly
by the quantum Rabi model~\cite{Rabi1936, Braak2011} which also takes into account the counterrotating terms describing processes where the number of excitations is no longer conserved. Reaching the regime of ultrastrong coupling paves the way for various applications and the study of interesting phenomena. For instance, it allows for the realization of ultrafast gates\cite{Romero2012} and provides deeper insight into Zeno physics~\cite{Lizuain2010} or photon transfer through cavity arrays~\cite{Felicetti2014}. Furthermore, a protocol allowing to simulate the regime of ultrastrong coupling with a standard circuit QED setup has been suggested~\cite{Ballester2012}. Such simulations can be used to interpret the results obtained in actual ultrastrong  coupling experiments.\\
In this work, we demonstrate physics beyond the Jaynes-Cummings model in a circuit QED architecture consisting of two coplanar stripline resonators and a superconducting flux qubit coupled galvanically to both of them. We discuss the resonant mode structure of this system and present a detailed analysis of the achieved high coupling strength. The multimode structure of our system provides an unambiguous spectroscopic proof for the breakdown of the Jaynes-Cummings approximation. Furthermore, we find that ultrastrong coupling of a qubit to a distributed resonator structure can be reached solely by the geometrical configuration of the latter without making use of additional inductive elements realized for example by Josephson junctions.

\section{Sample configuration and measurement setup}
\label{Sec:Setup}
\begin{figure}[t]
    \centering{\includegraphics{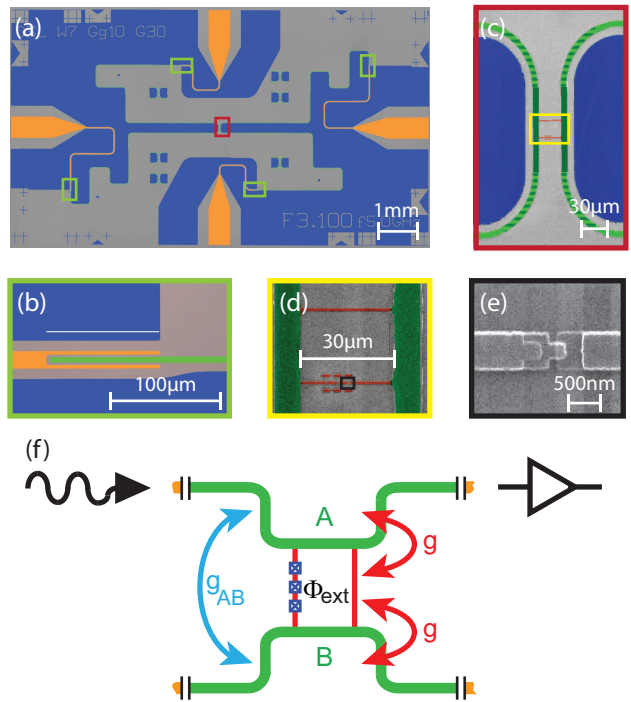}}
   \caption{Sample and measurement setup. (a) False-color image of the sample chip. Nb ground planes are shown in blue and feed lines in orange. The resonator signal lines reside along the ground plane edges. The green and red rectangles mark the areas shown on an enlarged scale in (b) and (c), respectively. (b) Coupling capacitor defining the resonators. (c) Resonator coupling area with signal lines (green) and flux qubit (red). Light/dark green stripes highlight Nb-Al overlap areas. The yellow rectangle marks the area shown in (d). (d) Flux qubit galvanically coupled to both resonators. The black rectangle marks the area shown in (e). (e) Al/$\rm{AlO_x}$/Al Josephson junction fabricated using shadow evaporation. (f) Sketch of the coupling mechanisms and measurement setup. The wiggly arrow symbolizes the input microwave line connected to resonator $A$ and the black triangle denotes the corresponding output line featuring microwave amplifiers. The crosses intersecting one qubit branch symbolize the three Josephson junctions.}
  \label{Baust_Fig1}
\end{figure}

Our sample is composed of two coplanar stripline resonators, $A$ and $B$, fabricated in Nb technology on a thermally oxidized Si substrate with fundamental mode frequencies $\omega_\mathrm{R}/2\pi\,{=}\,\SI{4.896}{\giga\hertz}$, cf.~Fig.~\ref{Baust_Fig1}(a) and Fig.~\ref{Baust_Fig1}(b). The detuning between the two resonators is found to be small and therefore disregarded. A superconducting persistent current flux qubit~\cite{Niemczyk2009} is coupled galvanically to the signal lines of both resonators at the position of the current antinodes of the lowest frequency modes as shown in Fig.~\ref{Baust_Fig1}(c)-(e). The flux qubit consists of a superconducting Al loop intersected by three Josephson junctions. Two of them have the critical current $I_\mathrm{c}$ and the phase drops across them are denoted by $\phi_1$ and $\phi_2$. The third one has a junction area smaller by a factor $\alpha\,{\simeq}\,0.7$. We mount the sample inside a gold-plated copper box attached to the base temperature stage of a dilution refrigerator stabilized at $\SI{45}{\milli\kelvin}$. The magnetic flux $\Phi_\mathrm{ext}$ applied to the qubit can be adjusted by means of a superconducting solenoid mounted on top of the sample box. Since some of the nomenclature used in the present work was introduced in previous work on this sample~\cite{Baust2014}, we briefly reiterate the main findings here. As discussed in Ref.~\cite{Baust2014}, the qubit can be used to tune and switch the coupling between the two resonators. In addition to the geometric coupling $g_\mathrm{AB}/2\pi\,{=}\,\SI{8.4}{\mega\hertz}$ there is the qubit mediated second-order dynamic coupling which depends on the magnetic flux applied to the qubit loop and on the qubit state. If the qubit is in the ground state, there exist certain flux values which we refer to as \emph{switch setting conditions} where the geometric coupling is (in the ideal case) fully compensated by the dynamical coupling such that the total coupling between the two resonators vanishes~\cite{Mariantoni2008}. Conversely, when the qubit is saturated by means of a strong excitation signal, the dynamical coupling is zero regardless of the flux applied to the qubit loop and the total coupling between the two resonators is given by $g_\mathrm{AB}$. The dependence of the dynamical coupling on the qubit state can be used to realize switchable coupling between the two resonators. As demonstrated in Ref.~\cite{Baust2014}, setting the flux operation point to a switch setting condition and applying a drive pulse to the qubit allows one to switch the coupling between the resonators $A$ and $B$ to the desired value between zero and $g_\mathrm{AB}$ depending on the drive pulse amplitude. This tunable coupler physics involves only two particular modes of the device. However, as we discuss in the following, the nature of the galvanic qubit-resonator coupling implies a more complex mode structure.

\section{Mode structure}
\label{Sec:Modestructure}
\begin{figure}[t]
    \centering{\includegraphics{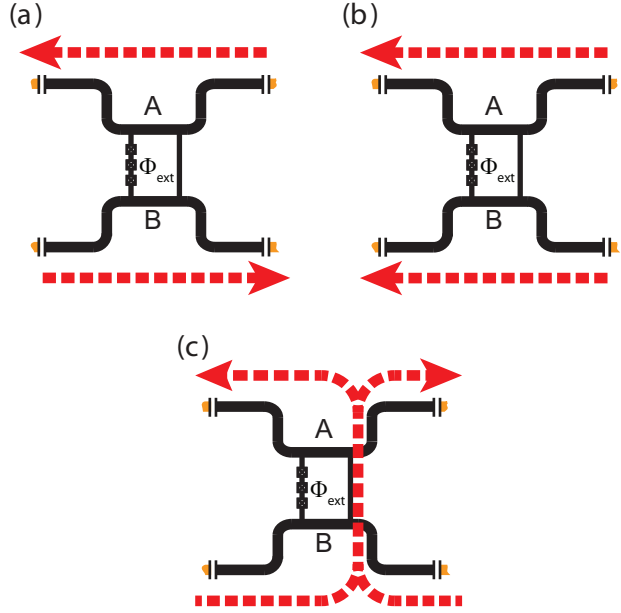}}
   \caption{Resonant modes of the galvanically coupled qubit-resonator system. The arrows indicate in-phase and out-of-phase oscillating currents. (a) Antiparallel mode. (b) Parallel mode. (c) Transverse mode.}
  \label{Baust_Fig2}
\end{figure}

\begin{figure}[t]
    \centering{\includegraphics{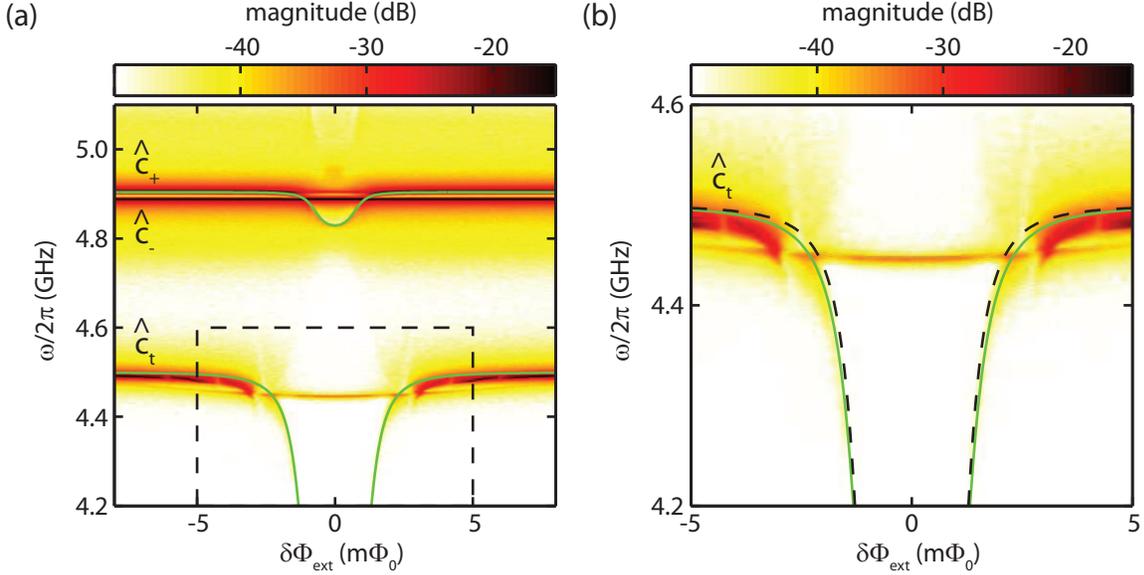}}
   \caption{(a) Transmission measured through resonator $A$ depending on the applied magnetic flux with the qubit in the ground state. Green line: Fit using the Hamiltonian of Eq.~(\ref{eq:fullHamilton}). The area shown in panel (b) is marked by the black rectangle. (b) Detail of (a). Solid green line: Fit using the Hamiltonian of Eq.~(\ref{eq:fullHamilton}). Dashed black line: Description within the Jaynes-Cummings model.}
  \label{Baust_Fig3}
\end{figure}

We first probe the coupled qubit-resonator system by measuring the transmission through resonator $A$ depending on the magnetic flux applied to the qubit loop, cf.~Fig.~\ref{Baust_Fig1}(f). For the measurement, the qubit is kept in the ground state and the input power is chosen such that the mean resonator population is approximately one photon on average. For coupled microwave resonators, we expect to observe two resonant modes corresponding to out-of-phase and in-phase oscillating currents in the two resonators, cf.~Fig.~\ref{Baust_Fig2}(a) and Fig.~\ref{Baust_Fig2}(b). Following the nomenclature in Ref.~\cite{Baust2014}, we refer to these modes as the antiparallel and parallel mode and assign to them the annihilation operators $\hat{c}_+$ and $\hat{c}_-$, respectively.
They can be identified in the spectroscopy data presented in Fig.~\ref{Baust_Fig3}(a). Far away from the qubit degeneracy point $\delta\Phi_\mathrm{ext}\,{\equiv}\,\Phi_\mathrm{ext}\,{-}\,\Phi_0/2\,{=}\,0$, where $\Phi_\mathrm{ext}$ is the external magnetic flux and $\Phi_0$ is the flux quantum, the dynamical coupling is negligible and the resonant frequencies of the antiparallel and parallel mode are given by $\omega_\mathrm{+}/2\pi\,{=}\,(\omega_\mathrm{R}\,{+}\,g_\mathrm{AB})/2\pi\,{=}\,\SI{4.904}{\giga\hertz}$ and $\omega_\mathrm{-}/2\pi\,{=}\,(\omega_\mathrm{R}\,{-}\,g_\mathrm{AB})/2\pi\,{=}\,\SI{4.888}{\giga\hertz}$.

However, the galvanic coupling of the qubit to both resonators gives rise to a third mode $\hat{c}_\mathrm{t}$ which we refer to as the `transverse mode'. It is identified as a parallel mode across the qubit as shown in Fig.~\ref{Baust_Fig2}(c). Far away from the qubit degeneracy point, its resonant frequency is found to be $\omega_\mathrm{t}/2\pi\,{=}\,\SI{4.508}{\giga\hertz}$. To explain the large frequency detuning between the transverse and the (anti)parallel mode, we assume that the inductance of the qubit has to be taken into account in order to correctly describe the frequency of the transverse mode. Following Refs.~\cite{Wallquist2006}~and~\cite{Sandberg2008}, we calculate the resonant frequency of the transverse mode to $\omega_\mathrm{t}\,{=}\,\omega_\mathrm{R}/(1\,{+}\,L_\mathrm{Q}/L_\mathrm{R})$, where $L_\mathrm{Q}$ is the inductance of the flux qubit and $L_\mathrm{R}$ is the inductance of a single resonator. The latter is given by $L_\mathrm{R}\,{=}\,Z/2\omega_\mathrm{R}\,{=}\,\SI{8.2}{\nano\henry}$, where $Z\,{=}\,80\,\Omega$ is the characteristic impedance of the resonator~\cite{gorur1999}.
The inductance of the flux qubit is given by $L_\mathrm{Q}\,{=}\,(\partial^2 U_\mathrm{Q}/\partial \Phi_\mathrm{ext}^2)^{-1}$, where $U_\mathrm{Q}\,{=}\,E_\mathrm{J}[2\,{+}\,\alpha\,{-}\,\cos\phi_1\,{-}\,\cos\phi_2\,{-}\,\alpha\cos(2\pi f\,{+}\,\phi_1\,{-}\,\phi_2)]$ is the flux qubit potential~\cite{Orlando1999}, $f\,{=}\,\Phi_\mathrm{ext}/\Phi_0$ is the frustration and $E_\mathrm{J}\,{=}\,\Phi_0 I_\mathrm{c} / 2\pi$ is the Josephson energy. Introducing $\phi_-\,{\equiv}\,(\phi_1-\phi_2)/2$, the inductance of the flux qubit reads as $L_\mathrm{Q}\,{=}\,\Phi_0/[2\pi\alpha I_\mathrm{c}\cos(2\pi f\,{+}\,2\phi_-)]$ with a minimum value of $L_\mathrm{Q}(f\,{=}\,0,\phi_-\,{=}\,0)\,{=}\,\Phi_0/2\pi\alpha I_\mathrm{c}\,{=}\,\SI{719}{\pico\henry}$, yielding a resonant frequency of $\omega_\mathrm{t,theo}/2\pi\,{=}\,\SI{4.501}{\giga\hertz}$. This value is in excellent agreement with the experimental value $\omega_\mathrm{t}/2\pi\,{=}\,\SI{4.508}{\giga\hertz}$ measured far away from the degeneracy point. To keep the theoretical modelling simple, in the following we assume a constant transverse mode frequency $\omega_\mathrm{t}$. That is, we assume that the experimentally observed flux dependence is solely due to the interaction with the flux qubit.

To gain further insight, we consider the Hamiltonian describing the coupling of the qubit to all resonant modes:
\begin{align}\label{eq:fullHamilton}
        \hat{H} &= \hat{H}_\mathrm{Q} + \sum_{\mathclap{\mathrm{\substack{n=\\\{+,-,t,3t,3+\}}}}}\hat{H}_n \nonumber\\
        &+ \hbar g\sqrt{2}\;\hat{\sigma}_z(\hat{c}_+^\dag\,+\,\hat{c}_+) \nonumber\\
        &+ \hbar g_\mathrm{t}\; \hat{\sigma}_z (\hat{c}_\mathrm{t}^\dag + \hat{c}_\mathrm{t})\nonumber\\
        &+ \hbar g_\mathrm{3t}\; \hat{\sigma}_z (\hat{c}^\dag_\mathrm{3t} + \hat{c}_\mathrm{3t})\nonumber\\
        &+ \hbar g_\mathrm{3+}\; \hat{\sigma}_z (\hat{c}^\dag_\mathrm{3+} + \hat{c}_\mathrm{3+}).
\end{align}

Here, $\hat{H}_\mathrm{Q}\,{\equiv}\,(\varepsilon/2)\hat{\sigma}_z\,{+}\,(\Delta/2)\hat{\sigma}_x$ is the qubit Hamiltonian and $\hat{H}_n\,{\equiv}\,\hbar\omega_n\hat{c}_n^\dag\hat{c}_n$ is the Hamiltonian describing the resonant mode $\hat{c}_n$. $\Delta$ is the qubit energy gap, $\epsilon(\Phi_\mathrm{ext})\,{=}\,2I_\mathrm{p}\delta\Phi_\mathrm{ext}$ denotes the qubit energy bias, and $I_\mathrm{p}\,{=}\,I_\mathrm{c}\sqrt{1-(2\alpha)^{-2}}$ the qubit persistent current. $\hat{\sigma}_x$ and $\hat{\sigma}_z$ are the Pauli operators.
As shown in Ref.~\cite{Baust2014}, the coupling of the qubit to the antiparallel mode is given by $g_+\,{=}\,\sqrt{2}g$ whereas there is virtually no coupling of the qubit to the parallel mode as the latter does not generate a magnetic field at the position of the qubit. To increase precision of our description, we also take into account the third harmonic of the $\hat{c}_\mathrm{t}$-mode (denoted by $\hat{c}_\mathrm{3t}$, located at $\omega_\mathrm{3t}/2\pi\,{=}\,\SI{13.1}{\giga\hertz}$) and the third harmonic of the $\hat{c}_+$-mode (denoted by $\hat{c}_\mathrm{3+}$, at $\omega_\mathrm{3+}/2\pi\,{=}\,\SI{14.3}{\giga\hertz}$). We do not consider the second harmonics since they exhibit current nodes at the qubit position and therefore do not couple to the qubit. The coupling strengths $g_\mathrm{3t}$ and $g_\mathrm{3+}$ are not considered as independent parameters, but are calculated via $g_\mathrm{3t}/2\pi\,{=}\,(g_\mathrm{t}/2\pi)\sqrt{\omega_\mathrm{3t}/\omega_\mathrm{t}}$ and $g_\mathrm{3+}/2\pi\,{=}\,(g_+/2\pi)\sqrt{\omega_\mathrm{3+}/\omega_{+}}$, taking into account the current distribution in the resonator. Fitting the Hamiltonian of Eq.~(\ref{eq:fullHamilton}) to our data (cf.~Fig.~\ref{Baust_Fig3}), the qubit energy gap is determined to $\Delta/h\,{=}\,\SI{3.55}{\giga\hertz}$ and the persistent current to $I_\mathrm{p}\,{=}\,\SI{458}{\nano\ampere}$. We find that the coupling strength between the qubit and each resonator is given by $g/2\pi\,{=}\,\SI{96.7}{\mega\hertz}$ and the coupling strength of the mode $\hat{c}_\mathrm{t}$ to the qubit is $g_\mathrm{t}/2\pi\,{=}\,\SI{775}{\mega\hertz}$ which is as high as $\SI{17.2}{\percent}$ of the respective mode frequency. Remarkably, the coupling strength even exceeds the relative coupling strengths observed in Ref.~\cite{niemczyk_circuit_2010} although the coupling is determined solely by the geometrical properties of the qubit arm and not by an additional inductive element such as a Josephson junction introduced in Ref.~\cite{niemczyk_circuit_2010} to enhance the coupling strength.
To understand the origin of the exceptionally large coupling strength, we assume that the coupling strength of the qubit to resonator $A$ and $B$, respectively, is determined by the shared arms between the qubit and the resonators $A$ and $B$, respectively. We further assume that the transverse mode current is flowing predominantly through the qubit arm without Josephson junctions as shown in Fig.~\ref{Baust_Fig2}(c). This assumption is well justified since the geometrical inductance of the qubit arm without Josephson junctions is much smaller than the total inductance of the branch containing the three Josephson junctions.

Following Ref.~\cite{Terman1945}, we can estimate the geometric inductance of the qubit branch connecting the two resonators $A$ and $B$ (length $\SI{30}{\micro\meter}$, width $\SI{0.5}{\micro\meter}$, thickness $\SI{0.1}{\micro\meter}$) at $\SI{31}{\pico\henry}$ which adds to the kinetic inductance~\cite{Schwarz2013} of approx.~$\SI{27}{\pico\henry}$, yielding a total inductance of the qubit branch $L_\mathrm{t}\,{=}\,\SI{58}{\pico\henry}$.
We further can estimate the inductance of the shared arms (length $\SI{20}{\micro\meter}$) between the resonators (total length $\SI{11.55}{\milli\meter}$) $A$ and $B$ and the qubit at $L_\mathrm{r}\,{=}\,L_\mathrm{R}\,{\cdot}\,\SI{20}{\micro\meter}/\SI{11.55}{\milli\meter}\,{=}\,\SI{14.2}{\pico\henry}$. 
The coupling strength $g_+$ between the antiparallel mode and the flux qubit is given by $\hbar g_+\,{=}\,2L_\mathrm{r} I_\mathrm{p} I_\mathrm{+}$ where $I_+\,{=}\,\sqrt{\hbar\omega_\mathrm{R}/2L_\mathrm{R}}$ is the vacuum current of the antiparallel mode~\cite{niemczyk_circuit_2010}. The total coupling strength $g_\mathrm{t}$ of the mode $\hat{c}_\mathrm{t}$ to the qubit is comprised of two contributions. The first one is the coupling mediated by the shared branch between qubit and resonator and the second one is the coupling mediated by the qubit branch connecting the two resonators. Therefore, we can calculate the coupling strength $\hbar g_\mathrm{t}\,{=}\,2L_\mathrm{r}I_\mathrm{p}I_\mathrm{t}\,{+}\,2L_\mathrm{t}I_\mathrm{p}I_\mathrm{t}$, where $I_\mathrm{t}$ is the vacuum current of the mode $\hat{c}_\mathrm{t}$. With these results, we estimate a ratio $g_\mathrm{t}/g_+\,{=}\,(\omega_\mathrm{t}/\omega_\mathrm{R})\cdot(2L_\mathrm{t}+2L_\mathrm{r})/2L_\mathrm{r}\approx 4.7$ in good agreement with the experimentally found ratio of $5.7$.

\section{Ultrastrong coupling}
In what follows, we discuss the theoretical framework needed to describe the interaction between the qubit and the multimode structure arising from our two-resonator circuit QED architecture. First, we rotate the Hamiltonian of Eq.~(\ref{eq:fullHamilton}) into the qubit eigenbasis using the transformations $\hat{\sigma}_\mathrm{z}\,{\rightarrow}\,\cos\theta\hat{\sigma}_\mathrm{z}\,{-}\,\sin\theta\hat{\sigma}_\mathrm{x}$ and $\hat{\sigma}_\mathrm{x}\,{\rightarrow}\,\sin\theta\hat{\sigma}_\mathrm{z}\,{+}\,\cos\theta\hat{\sigma}_\mathrm{x}$,
where $\sin\theta\,{\equiv}\,\Delta/\hbar\omega_\mathrm{q}$ and $\cos\theta\,{\equiv}\,\epsilon/\hbar\omega_\mathrm{q}$ and $\hbar\omega_\mathrm{q}\,{=}\,\sqrt{\Delta^2+\epsilon^2}$ is the flux-dependent qubit transition energy. In the qubit eigenbasis, the Hamiltonian reads
\begin{equation}\label{eq:rotHam}
  \hat{H}^{*}\,{=}\,\hat{H}_\mathrm{Q}^{*}\,{+}\,\sum_{\mathclap{\mathrm{\substack{n=\\\{t,+,3t,3+\}}}}} \left[\hat{H}_n\,{+}\,\hbar g_n\left(\hat{c}^\dagger_n\,{+}\,\hat{c}_n\right)(\cos\theta\hat{\sigma}_\mathrm{z}\,{-}\,\sin\theta\hat{\sigma}_\mathrm{x})\right]
\end{equation}
with $\hat{H}_\mathrm{Q}^{*}\,{=}\,\frac{\hbar\omega_\mathrm{q}}{2}\hat{\sigma}_\mathrm{z}$. At $\Phi_\mathrm{ext}\,{=}\,\Phi_0/2$, the Hamiltonian of Eq.~(\ref{eq:rotHam}) represents a multimode quantum Rabi model. We note that we drop the $\hat{c}_-$-mode since it does not couple to the qubit.
Defining the qubit state raising and lowering operators $\hat{\sigma}_\pm\,{\equiv}\,(\hat{\sigma}_x\,{\pm}\,i\hat{\sigma}_y)/2$, we find that the Hamiltonian of Eq.~(\ref{eq:rotHam}) explicitely contains counterrotating terms of the form $\hat{c}^\dagger_n\hat{\sigma}_+$ and $\hat{c}_n\hat{\sigma}_-$. For $g_n\,{\ll}\,\omega_n$, a rotating wave approximation reduces the Hamiltonian of Eq.~(\ref{eq:rotHam}) to the well known multimode Jaynes-Cummings Hamiltonian for arbitrary $\Phi_\mathrm{ext}$. Following Ref.~\cite{niemczyk_circuit_2010}, the regime of \emph{ultrastrong coupling} is reached when the interaction between the qubit and one or multiple modes can be described by the quantum Rabi model, but qualitative deviations from the Jaynes-Cummings model are observed. Despite these deviations, the system dynamics still reflects the intuition of several distinct, but coupled systems exchanging excitations. This intuition breaks down completely in the \emph{deep strong coupling regime}~\cite{Casanova2010}, where $g\,{\gtrsim}\,\omega$ and the dynamics of the system is characterized by the emergence of two parity chains.

\section{Breakdown of the Jaynes-Cummings model}

\begin{figure*}[htb]
\centering{\includegraphics{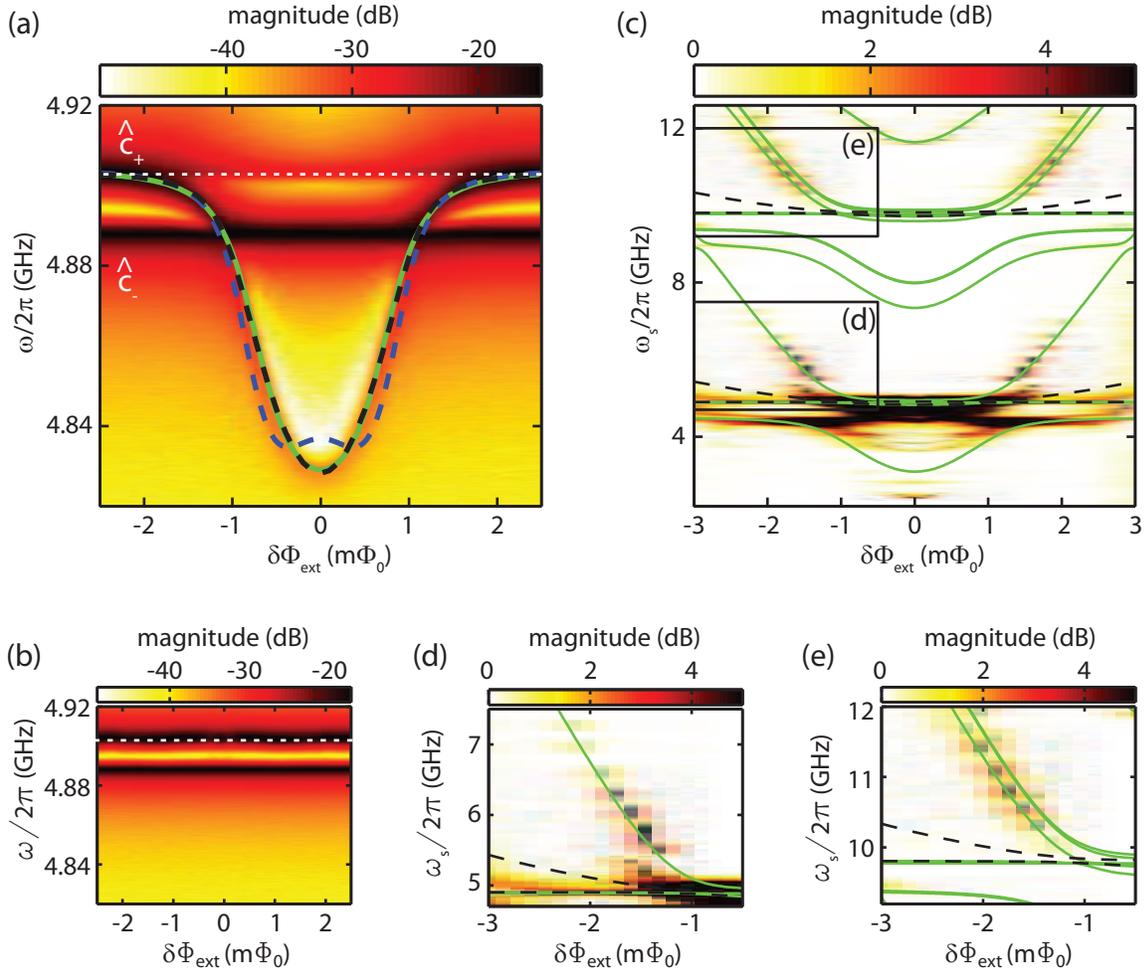}}
   \caption{Breakdown of the Jaynes-Cummings model. (a) Transmission measured through resonator $A$ depending on the magnetic flux applied to the flux qubit (detail from Fig.~\ref{Baust_Fig2}) with the qubit in the ground state. Green line: Fit of the full Hamiltonian of Eq.~(\ref{eq:fullHamilton}). Blue dashed line: Prediction by the Jaynes-Cummings model. Black dashed line: Fit to the Jaynes-Cummings model neglecting the transverse mode. White dashed line: Measurement frequency for two-tone spectroscopy. (b) Same as (a), qubit driven with strong excitation signal. (c) Two-tone spectroscopy. Green lines: Fit of the spectroscopy data to the full Hamiltonian (\ref{eq:fullHamilton}). Black dashed lines: Description within the Jaynes-Cummings model. (d, e) Details from (c).}
  \label{Baust_Fig4}
\end{figure*}

Next, we analyze whether our multipartite circuit QED setup comprised of a flux qubit and two galvanically coupled resonators is consistent with the Jaynes-Cummings model or whether it has to be treated within the more general Rabi model. First, we assume that the Rabi model represents a valid theoretical model for our setup and fit the Hamiltonian of Eq.~(\ref{eq:rotHam}) to our spectroscopy data. As shown by Fig.~\ref{Baust_Fig4}(a) theory and experimental data agree very well for the $\hat{c}_+$-mode. However, if we drop the counterrotating terms without making a new fit, we find a pronounced qualitative deviation between our experimental data and the Jaynes-Cummings model prediction. The observed deviations are in agreement with the observation of the Bloch-Siegert shift in a system comprised of a flux qubit coupled ultrastrongly to an $LC$-resonator~\cite{Forndiaz2010}. Figure~\ref{Baust_Fig3}(b) shows the fit of the full Hamiltonian to our spectroscopy data for the transverse mode $\hat{c}_\mathrm{t}$ and the corresponding description within the Jaynes-Cummings model. Even if a small quantitative difference can be observed, there is no qualitative difference between the two models. This can be understood considering the fact that the Bloch-Siegert shift is proportional to $g^2\sin^2\theta/(\omega_\mathrm{q}+\omega_\mathrm{r})$ and, hence, is most prominent near the qubit degeneracy point. However, the pronounced qualitative deviation between Rabi and Jaynes-Cummings model for the $\hat{c}_+$-mode (cf.~Fig.~\ref{Baust_Fig4}(a)) indicates that the rotating wave approximation is no longer valid presuming that the Rabi model correctly describes our experimental findings.
\\One also may argue that the transverse mode $\hat{c}_\mathrm{t}$ is an independent phenomenon. To check whether the Jaynes-Cummings model then eventually can correctly describe our data, we omit the transverse mode in the Hamiltonian of Eq.~(\ref{eq:fullHamilton}) and subsequently apply a rotating wave approximation to the remainder of Eq.~(\ref{eq:fullHamilton}).
As shown by Fig.~\ref{Baust_Fig4}(a), the resulting Hamiltonian
\begin{align}\label{eq:remainHamilton}
        \hat{H} &= \hat{H}_\mathrm{Q}^{*} + \sum_{\mathclap{\mathrm{n=\{+,3+\}}}}\hat{H}_n\nonumber\\
        &+ \hbar g_+(\hat{c}_+^\dag \hat{\sigma}_- + \hat{c}_+\hat{\sigma}_+) \nonumber\\
        &+ \hbar g_\mathrm{3+}(\hat{c}_{3+}^\dag \hat{\sigma}_- + \hat{c}_{3+}\hat{\sigma}_+).
\end{align}
can obviously be successfully used to fit our data. However, even if the fit looks nice, this ansatz yields qubit parameters deviating strongly from the qubit parameters given in Sec.~\ref{Sec:Modestructure}, where we performed the fit using the Hamiltonian of Eq.~(\ref{eq:rotHam}). In order to verify which of the two parameter sets is incorrect, we make use of the fact that our measurement setup does not only provide access to the eigenmodes of the coupled qubit-resonator system, but also allows to perform spectroscopy of the qubit using a two-tone spectroscopy experiment. To this end, we record the transmission through resonator $A$ at the frequency of $\omega_+/2\pi\,{=}\,\SI{4.904}{\giga\hertz}$. When the qubit is far detuned, this corresponds to the resonant frequency of the $\hat{c}_+$-mode. In addition, a second microwave tone, the \emph{spectroscopy tone}, with variable frequency $\omega_\mathrm{s}$ is applied to the coupled qubit-resonator system via the input port of resonator $B$. When the qubit is in the ground state, the measured transmission as a function of the magnetic flux applied to the qubit loop corresponds to a cut through Fig.~\ref{Baust_Fig4}(a) along $\omega_+/2\pi$ as highlighted by the white dashed line. When the qubit is saturated by means of the spectroscopy tone, the qubit state is described by the density matrix $\rho_\mathrm{M}\,{=}\,\frac{1}{2}(|g\rangle\langle g|\,{+}\,|e\rangle\langle e|)$ and the transmission spectrum turns into the one shown in Fig.~\ref{Baust_Fig4}(b). Evidently, the transmission magnitude at $\omega_+/2\pi$ increases near the degeneracy point when the qubit is driven. Using this protocol, we record the change in resonator transmission as a function of the spectroscopy tone frequency $\omega_\mathrm{s}$ and the applied magnetic flux, cf.~Figs.~\ref{Baust_Fig4}(c)-(e).
We compare the measured data to the energy level spectrum of the Hamiltonian of Eq.~(\ref{eq:rotHam}) by calculating the energy differences between the ground state and the $15$ lowest energy levels. As can be seen, there is very good agreement between our two-tone spectroscopy data and their description within the full Hamiltonian of Eq.~(\ref{eq:rotHam}). However, the energy level spectrum calculated from the qubit parameters found by a fit of the $\hat{c}_+$-mode within the Jaynes-Cummings approximation clearly deviates from the two-tone spectroscopy data. In other words, treating the mode $\hat{c}_\mathrm{t}$ independently of the mode $\hat{c}_+$ clearly does not allow us to correctly describe our experimental data within the Jaynes-Cummings model.\\
Finally, we compare our  findings to previous work on ultrastrong coupling in superconducting circuits. In the present sample, the access to both resonator and qubit spectroscopy data allows us to rigorously rule out the validity of the Jaynes-Cummings model without having to assume the validity of the Rabi model.
Hence, our analysis goes beyond the treatment presented in Ref.~\cite{Forndiaz2010}. In addition, the present work is markedly different from the approach used in Ref.~\cite{niemczyk_circuit_2010}. There, it was shown that in a multimode system the number of excitations is no longer preserved in the ultrastrong coupling regime. Despite this difference, it appears that physics beyond the Jaynes-Cummings model in circuit QED is favourably demonstrated by analyzing the complex mode structure of multipartite setups.

\section{Conclusions}
In conclusion, we demonstrate the breakdown of the Jaynes-Cummings model in a system comprised of two coplanar stripline resonators and a persistent current flux qubit coupled galvanically to both of them. We analyze the complex mode structure and find that the coupling of one resonant mode to the qubit reaches $17\%$ of the mode frequency, exceeding that of previous circuit QED experiments~\cite{niemczyk_circuit_2010, Forndiaz2010}. We show that both the mode frequency and the coupling strength are in good agreement with theoretical calculations based on the quantum Rabi model. Analyzing the resonator and qubit spectroscopy data clearly shows that the Jaynes-Cummings model does no longer provide an appropriate description of the observed behaviour, confirming that our circuit QED setup is in the regime of ultrastrong coupling. In our sample, a remarkably large coupling strength is reached without utilizing the inductance of an additional Josephson junction. As a future perspective, combining different methods for enhancing the coupling strength may provide access to the regime of deep strong coupling~\cite{Casanova2010}, giving exerimental insight into a completely novel regime of light-matter interaction.

\section*{Acknowledgements}
This work is supported by the German Research Foundation through SFB 631, EU projects CCQED,
PROMISCE and SCALEQIT, Spanish MINECO projects FIS2012-33022 and FIS2012-36673-C03-02, UPV/EHU
UFI 11/55, UPV/EHU PhD Grant and Basque Government IT472-10.

\section*{Appendix}
\begin{figure}[htb]
    \centering{\includegraphics{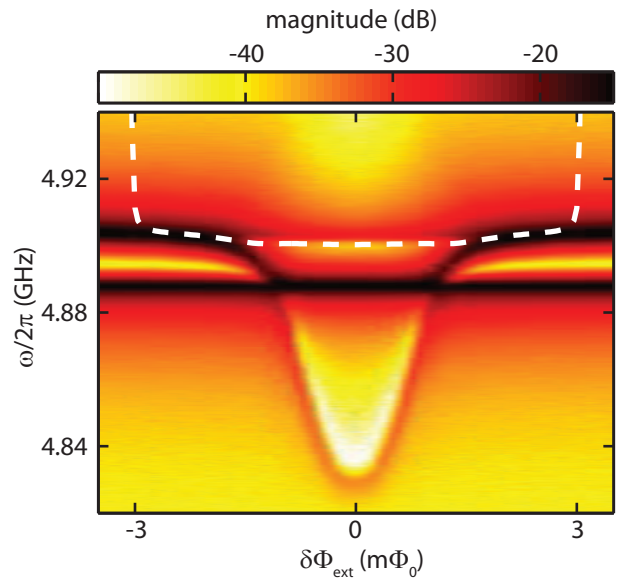}}
   \caption{Transmission measured through resonator $A$ depending on the magnetic flux applied to the flux qubit. White dashed line: Transition between eigenstates of the Hamiltonian of Eq.~(\ref{eq:fullHamilton}) as described in the main text.}
  \label{Baust_Fig5}
\end{figure}

In this section, we briefly discuss the origin of the resonant structure which is visible near the degeneracy point at the frequency of $\SI{4.904}{\giga\hertz}$, cf.~Fig.~\ref{Baust_Fig5}. We find good agreement between this resonant structure and the transition between the eigenstates corresponding to the second and the sixth lowest eigenenergies of the Hamiltonian of Eq.~(\ref{eq:fullHamilton}).
Compared to the antiparallel mode, the additional resonant structure is suppressed by approx.~$14\,\mathrm{dB}$. Hence, this structure might arise from a small finite population of the second energy level due to the finite sample temperature of $\SI{45}{\milli\kelvin}$ and the very large coupling strength of the qubit to the transverse mode. This interpretation is also in agreement with a similar resonant structure observed in Ref.~\cite{Forndiaz2010}.

\section*{References}
\bibliographystyle{iopart-num}
\bibliography{Bibliography_NJP_V1}

\end{document}